# Localização 3D em sistemas RFID com leitor móvel

Eric S. Ferraz[1], Celso B. Carvalho[1]

[1]Faculdade de Tecnologia – Universidade Federal do Amazonas (UFAM)
Manaus – AM – Brasil

ericferraz@gmail.com, ccarvalho_@ufam.edu.br

***Abstract.*** *RFID has been widely used in applications for indoor objects location. This article proposes a 3D location algorithm based on a system that uses a mobile reader and a reference matrix of real (passive) and virtual tags. The proposed algorithm compares the RSSI of the target tag with the RSSI of the tags of the reference matrix and defines the estimated position. Preliminary results show that the location scheme may be promising for indoor location of tags.*

***Resumo.*** *RFID vem sendo bastante utilizada em aplicações para localização de objetos em ambientes internos. Este artigo propõe um algoritmo de localização 3D baseado em um sistema que utiliza um leitor móvel e uma matriz de referência de etiquetas passivas e virtuais. O algoritmo proposto compara o RSSI da etiqueta alvo com o RSSI das etiquetas da matriz de referência e define a posição estimada. Resultados preliminares mostram que o esquema de localização poderá ser promissor na localização indoor de etiquetas.*

## 1. Introdução

O RFID (*Radio Frequency Identification*) é uma forma mais eficiente de identificar objetos que a convencional técnica de código de barras e vem se tornando popular em aplicações de vigilância, controle de acesso ou rastreamento de movimento [Finkenzeller 2010]. Um esquema de localização baseado em RFID atribui etiquetas a objetos e localiza estes objetos por meio da comunicação entre as etiquetas e os leitores fixados em locais conhecidos. LANDMARC [Ni, Liu, Lau and Patil 2004] é um dos trabalhos pioneiros no emprego da intensidade de sinais para localização interna, utiliza etiquetas de referência e pré-define o mapa de localização das etiquetas de referência para facilitar o rastreamento. Mais tarde VIRE [Zhao, Liu and Ni 2007] propõe a utilização de etiquetas de referência virtuais de forma a reduzir a distância entre etiquetas na grade de referência. Han, et al. [Han, Zhao, Cheng, and Wong and Wong 2012] estende a solução proposta em VIRE para o plano 3D, através da utilização do RSSI (*Received Signal Strength Indicator*) das etiquetas reais por meio de interpolação linear, encontrando o valor de RSSI das etiquetas virtuais. O sistema proposto por Wang, et al. [Wang, Wu, Shi and Zhang 2016] utiliza otimização por enxame de partículas (*Particle Swarm Optimization* - PSO) e filtros gaussianos para redução do erro de localização.

Após análise dos artigos de referência verificou-se todos utilizam leitores fixos (quanto maior a quantidade de leitores maior a acurácia). O aumento seja de leitores ou de etiquetas encarece o projeto, e por este motivo propomos a utilização de um único leitor móvel aliado a uma matriz de etiquetas de referência e etiquetas virtuais. Desta forma reduzimos os custos de implementação do sistema e conseguimos realizar a localização de etiquetas com apenas um leitor. A avaliação de desempenho do algoritmos de localização, realizado por simulação, apresenta um erro médio de 1.9 m.

Para tratar do assunto do tema localização 3D em sistemas RFID, na seção 2 apresentamos a nossa proposta e resultados preliminares e na seção 3 as conclusões e trabalhos futuros.

## 2. Proposta e Resultados Preliminares

De forma a validar o algoritmo proposto neste trabalho, optou-se pela simulação utilizando o software Matlab no qual desenvolvemos nosso próprio simulador. Implementamos uma matriz 3D com 3m de largura, 3m de comprimento e 4m de altura (simulando um cômodo de uma residência) com etiquetas de referência posicionadas estrategicamente nos oito vértices do ambiente e um leitor na posição "zero" (0,0,0). A potência refletida por cada etiqueta interrogada pelo leitor é dada pela equação de Friis:

$$P_{rx} = \frac{P_{tx} \cdot k \cdot G_{leitor}^{2} \cdot G_{etiqueta}^{2} \cdot \lambda^{4}}{(4 \cdot \pi \cdot d)^{4}} \quad (2.1)$$

onde: $P_{tx}$ = 30 dBm é o valor da potência transmitida pelo leitor, $G_{leitor}$ = 1 é o ganho da antena do leitor, $G_{etiqueta}$ = 1 é o ganho da antena da etiqueta, $\lambda$ = 0,3125 é o comprimento de onda na frequência de 960 MHz, $k$ = 0,33 [Lucena Filho, W. D. C. 2015] é a perda de transmissão *backscatered* e $d$ é a distância entre o leitor e a etiqueta dada pela equação:

$$d = \sqrt{\left((X_{ref} - X_{leitor})^{2} + (Y_{ref} - Y_{leitor})^{2} + (Z_{ref} - Z_{leitor})^{2}\right)} \quad (2.2)$$

onde: $X_{ref}$, $Y_{ref}$ e $Z_{ref}$ são as coordenadas das etiquetas de referências e $X_{leitor}$, $Y_{leitor}$ e $Z_{leitor}$ são as coordenadas do leitor.

A Figura 1 mostra a estratégia utilizada para posicionamento das etiquetas virtuais, $n$ etiquetas virtuais podem ser posicionadas equidistantes entre duas etiquetas de referência do ambiente simulado, esta estratégia visa diminuir o erro da matriz de referência reduzindo a distância entre as etiquetas da matriz de referência.

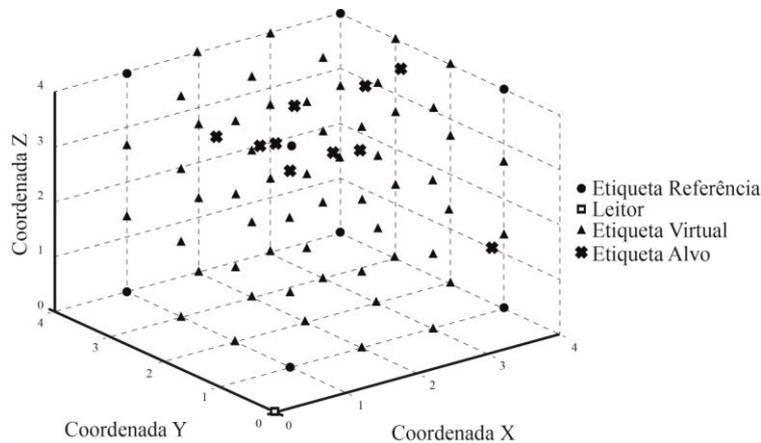

**Figura 1. Posicionamento das etiquetas no ambiente simulado**

A ideia desta abordagem é estimar a potência refletida pelas etiquetas virtuais utilizando interpolação linear e tendo como referência os valores de RSSI das etiquetas de referência. Desta forma, quando um objeto a ser monitorado, estando com uma etiqueta RFID (etiqueta alvo), posicionado aleatoriamente no ambiente simulado refletir para o leitor um valor de potência (RSSI), este pode ser comparado ao RSSI das etiquetas virtuais e/ou de referências a fim de se determinar o posicionamento do objeto monitorado.

O Algoritmo implementado para localização das etiquetas alvo é descrito abaixo:

---

**Algoritmo 1**: Algoritmo para localização de etiquetas RFID
**Entrada:** *Ptx_leitor*, *Frequency*, *Reader_Position*, *Matriz_Ref (RSSI, Positions)*
**Entrada:** *posições*, *RealTags_Position*, *count*
**Saída:** *Posição_Estimada, erro*
1: *Ptx_leitor* ← 30*dBm*; *Frequency* ← 960*MHz*; *Reader_Position* ← (0,0,0);
2: *Matriz_Ref (RSSI, Positions)* ← inf. de RSSI e Posições da matriz de referência;
3: *posições* ← 10; , *RealTags_Position* ← Posições rand. de coordenadas (*x, y, z*);
4: *count* ← 1;
5: **enquanto** *Reader_Position(1)* ≤ *posições* **faça**
6:   *Distancia_realtag* ← Distância [*Reader_Position x RealTags_Position*];
7:   *RSSI_realtag* ← Aplicação da equação de Friis em *Distancia_realtag*;
8:   *Menor_dif(i)* ← Compara *RSSI_realtag* com RSSI da *Matriz_Ref*;
9:   *Pos_tagmenordif(i)* ← Recebe a pos. da etiq. com < valor de diferença;
10:   *Reader_Position(1)* ← *Reader_Position(1)* + 1;
11:   *count* ← *count* + 1;
12: **fim enquanto**
13: *Posição_Estimada* ← Posição com maior incidência em *Pos_tagmenordif*;
14: *erro* ← Distancia [*Posição_Estimada x RealTags_Position*];

---

**Algoritmo 1. Algoritmo proposto para localização de etiquetas**

O algoritmo proposto gera 2 saídas:

- *Posição_Estimada*: Posição estimada da etiqueta alvo;

- *erro*: Distância encontrada entre a posição estimada e a real posição da etiqueta alvo, calculado através da Equação 2.2

As linhas de 1 a 4 são para inicialização das variáveis de entrada. As linhas de 5 a 12 são responsáveis pela identificação da posição da etiqueta alvo na matriz de referência. Esta posição é determinada nas linhas 8 e 9, sendo definida como a posição da etiqueta (de referência ou virtual) que possui a menor diferença no valor de RSSI quando comparado ao da etiqueta alvo. Já as linhas 13 e 14 são as saídas do algoritmo.

Foram executadas simulações alterando-se o valor de *n* (quantidade de etiquetas virtuais posicionadas entre duas etiquetas de referência). Os resultados da média do erro para cada valor de *n* são apresentados na Figura 2.

Expandindo o erro nas 3 coordenadas (*x, y, z*) temos:

erro eixo *x* = 0,17m; erro eixo *y* = 1,17; erro eixo *z* = 1,46m

Verificamos que o erro médio no eixo em que o leitor móvel se movimenta (eixo *x*) é inferior a 0.2m. Utilizando a mobilidade do leitor pretendemos efetuar medições de RSSI acrescentando movimento do leitor móvel nos eixos y e z de forma a reduzir o erro médio total.

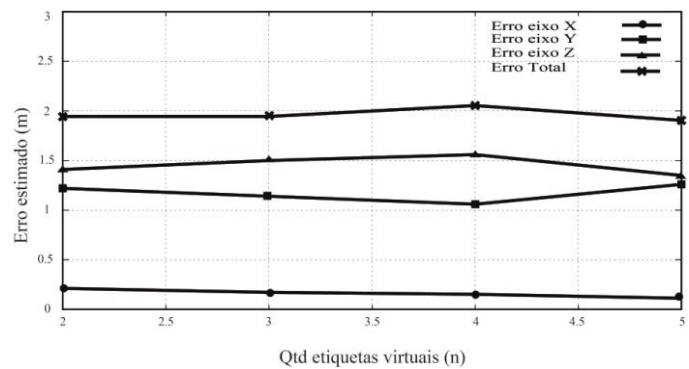

**Figura 2. Erros médios encontrados**

## 3. Conclusões e trabalhos futuros

Neste artigo propomos um mecanismo de localização de etiquetas RFID em ambientes 3D utilizando um único leitor RFID móvel. Tal cenário não tem sido verificado em outros trabalhos da literatura. O algoritmo de localização, ainda possui erro médio elevado na localização de etiquetas. Após uma análise dos resultados iniciais verificamos que o erro médio no eixo x (eixo no qual o leitor esta se movimentando nas simulações) é inferior a 0.2m. Na continuação da pesquisa, pretendemos efetuar a adaptação do algoritmo para cenários com variação de RSSI, utilização de filtros de Kalman para redução do erro da posição estimada e introdução de um controle de potência para redução da energia total utilizada.

## 6. Referências